\documentclass[a4paper]{jpconf}
\usepackage{graphicx}
\usepackage{epsf}

\newcommand{\seven}{$\sqrt{s}=7$ TeV}

\newcommand{\twosevensix}{$\sqrt{s_{NN}}=2.76$ TeV}

\begin{document}
\title{Strange and Multi-Strange Particle Production in ALICE}

\author{D.D. Chinellato$^{1}$ for the ALICE Collaboration}

\address{$^{1}$ University of Houston, Physics Department, 617 Science \& Research Building 1, Houston, TX 77204}

\ead{david.dobrigkeit.chinellato@cern.ch}

\begin{abstract}
The production of strange and multi-strange hadrons in proton-proton (pp) and lead-lead (Pb-Pb) collisions is studied with the ALICE experiment at the CERN LHC. These particles are reconstructed via their weak decay topologies, exploiting the tracking and particle  identification capabilities of ALICE. Measurements of central rapidity yields of $\Lambda$, $\Xi^{-}$ and $\Omega^{-}$ baryons, their antiparticles and ${\mathrm K}^{0}_{S}$ mesons are presented as a function of transverse momentum for Pb-Pb collisions at $\sqrt{s_{NN}}=2.76$ TeV. They are compared to those observed in pp collisions as well as to results from lower energy nucleus-nucleus measurements.
\end{abstract}

\section{Introduction}

The study of strangeness production in both proton-proton (pp) and lead-lead (Pb-Pb) collisions contributes to the understanding of particle production and hadrochemistry in general. This is not only due to the fact that there is no net strangeness content in the colliding systems but also because strange quarks are the lightest and most abundantly produced among the higher generation quarks. 

In proton-proton collisions, the strange hadron measurements serve not only as a benchmark for the heavy-ion collision measurements but may also offer an insight into the different production mechanisms at play. At low transverse momenta ($\ensuremath{p_{\mathrm{T}}}$), soft interactions are expected to be the dominant source of particle production, while at high $\ensuremath{p_{\mathrm{T}}}$ it is expected that interactions with high transferred momenta, i.e. those that can be described well with perturbative QCD (pQCD), take over as the most significant production mechanism. 

In heavy-ion collisions, it has been proposed that strange quarks may be produced thermally in the high energy density medium formed \cite{RafelskyStranEnh}. In this scenario, the final yields would exhibit a \emph{strangeness enhancement}. This is defined as a higher yield of strange and multi-strange hadrons in Pb-Pb collisions when compared to production rates measured in proton-proton collisions with an appropriate scaling for the volume, such as the number of participant nucleons ($N_{\mathrm{part}}$).

\section{Data Analysis}

Strange and multi-strange hadrons are reconstructed by their weak decay topologies in a sample of about $130\times10^{6}$ pp collisions at \seven\ and about $30\times10^6$ Pb-Pb collisions at \twosevensix, both collected in 2010. ${\mathrm K}^{0}_{S}$, $\Lambda$ and $\bar{\Lambda}$ are reconstructed by combining two charged tracks into V-shaped decays (the so-called V0 topology) while $\Xi^{-}$, $\bar{\Xi}^{+}$, $\Omega^{-}$ and $\bar{\Omega}^{+}$ are reconstructed by further combining one $\Lambda$ or $\bar{\Lambda}$ candidate with a third track (the so-called cascade decay topology). The V0 and cascade decay candidates are formed only if the tracks involved satisfy certain geometrical restrictions to ensure consistency with the expected topology. These initial selection criteria are thus called topological selections. Furthermore, the energy loss for ionization of the TPC gas is used as a particle identification criterion for all daughter tracks, reducing background substantially. An upper limit on the actual lifetime selection is also imposed in terms of the PDG $c\tau$ of each particle for the V0 analysis and serves to remove secondaries from material interactions. In addition, a competing decay rejection in Armenteros-Podolanski parameter space is applied for selecting ${\mathrm K}^{0}_{S}$ candidates in the Pb-Pb analysis, where only candidates satisfying $|\alpha_{arm}| < 5~\ensuremath{p_{\mathrm{T}}}^{arm}$ are accepted into the ${\mathrm K}^{0}_{S}$ analysis. In this equation, $\alpha_{arm}=(p^{+}_{||}-p^{-}_{||})/(p^{+}_{||}+p^{-}_{||})$, where $p_{||}$ is the momentum component of the positive or negative daughter particle parallel to the momentum of the parent, and $\ensuremath{p_{\mathrm{T}}}^{arm}$ is the momentum component perpendicular to the parent momentum given in {\mbox{\rm GeV$\kern-0.15em /\kern-0.12em c$}}. This selection has been shown to consistently remove background without creating any false peaks and acts to select decays that are more symmetric in phase space. Finally, in the case of multi-strange baryons, candidates with an invariant mass close to the $\Xi$ mass under the appropriate mass hypothesis are rejected for the $\Omega$ analysis.

Signal extraction is performed by sampling the background in areas adjacent to the invariant mass peaks and subtracting this background from the peak region. The background areas are $\ensuremath{p_{\mathrm{T}}}$ dependent to account for the variation of resolution with momentum. For the $\Lambda$ and $\bar{\Lambda}$ analysis, we subtract the contribution to the $\Lambda$ signal coming from $\Xi$ decays by first computing a feeddown matrix that relates the $\Xi$ production at a given momentum range to the secondary $\Lambda$ detection at any $\Lambda$ $\ensuremath{p_{\mathrm{T}}}$. Once this matrix is calculated from Monte Carlo simulations using the same selections as in the real data analysis, we use a real measurement as weight to obtain the $\Lambda$ signal that needs to be subtracted from the measurement. Feeddown from neutral $\Xi$  is roughly equal to the one from charged $\Xi$; together they account for about 20\% of the raw measured $\Lambda$ signal in both pp and Pb-Pb collisions. 

In order to compute detection efficiency, we use PYTHIA \cite{pythia6.4} (for pp) or HIJING \cite{HijingRef} (for Pb-Pb) event generators and subsequently propagate the produced particles through the full ALICE geometry with GEANT3 \cite{Geant3Ref, Geant3Ref2}. When computing this efficiency for Pb-Pb collisions and also for the multi-strange in pp, the Monte Carlo productions were enriched with injected particles in order to reduce the statistics required for proper determination of detection efficiencies.

Systematic uncertainties are evaluated point-by-point for the topological selections, signal extraction and track quality cuts. Other sources of systematic uncertainties such as normalization, feeddown correction, material budget, GEANT3 (anti-)proton cross-section, TPC particle identification, proper lifetime selection and competing decay rejection are taken to be $\ensuremath{p_{\mathrm{T}}}$ independent.

\section{Measurements in Proton-Proton Collisions}
\begin{figure}
\centerline{\includegraphics[width=3.45in]{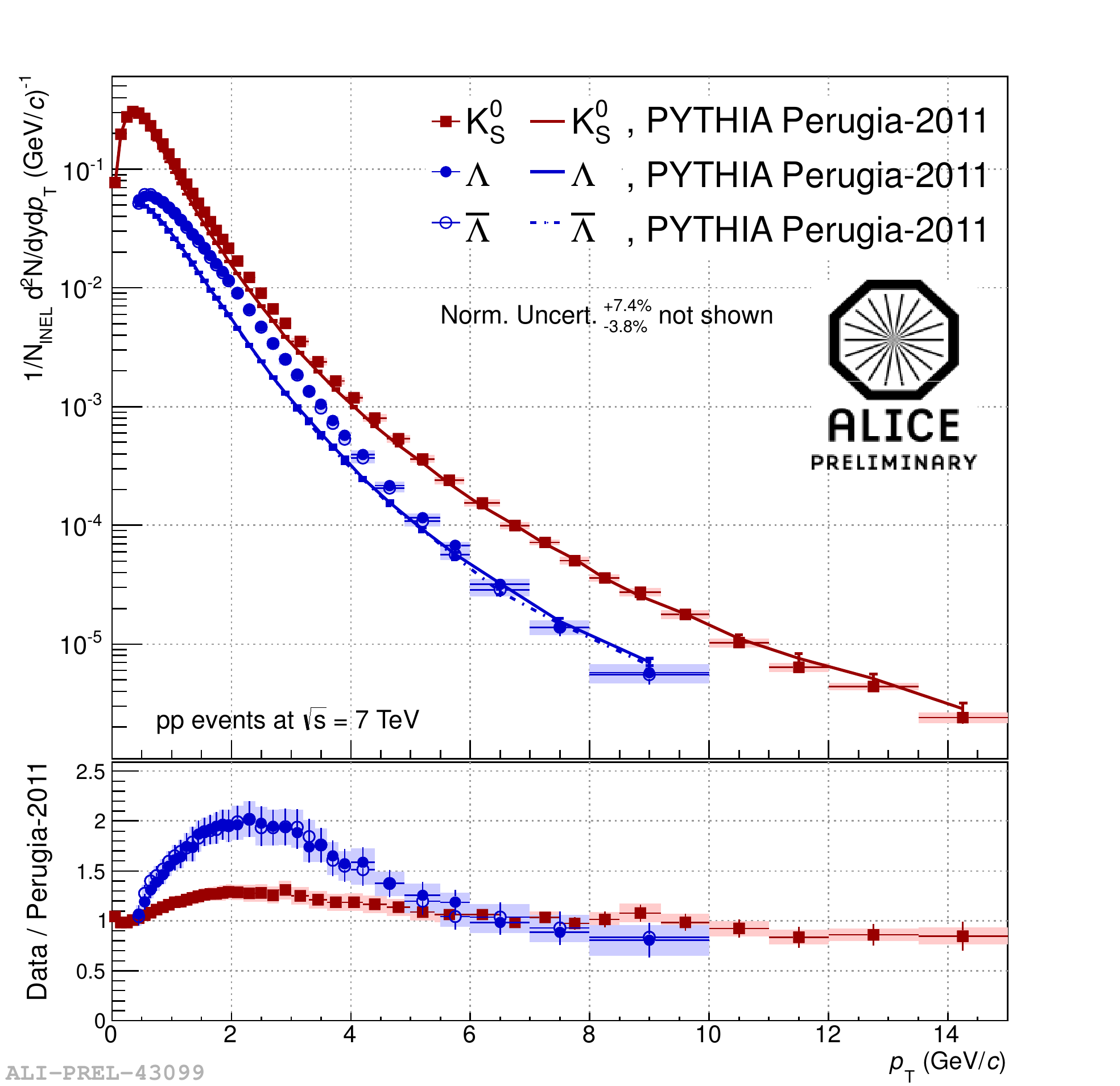}}
\caption{\label{ActualSpectra}  $\Lambda$, $\bar{\Lambda}$ and ${\mathrm K}^{0}_{S}$ transverse momentum spectra in pp collisions at \seven\  compared to Pythia P2011 predictions.}
\end{figure}

The ${\mathrm K}^{0}_{S}$, $\Lambda$ and $\bar{\Lambda}$ efficiency- and feeddown-corrected spectra in pp collisions at $\sqrt{s}$ = 7 TeV are shown in Fig. \ref{ActualSpectra}. The $\bar{\Lambda}/\Lambda$ ratio is compatible with unity throughout the measured transverse momentum range of 0.4~-~10.0~{\mbox{\rm GeV$\kern-0.15em /\kern-0.12em c$}}. For the extraction of the integrated production rates, we fit the spectra with Tsallis functions, as done previously in lower energy pp strangeness analyses in ALICE~\cite{ALICE_str_pp}. While a 15\% extrapolation at low ${\ensuremath{p_{\mathrm{T}}}}$ is required for the computation of the $\Lambda$ yields, only a negligible extrapolation is needed for the ${\mathrm K}^{0}_{S}$ yield.

The total production rates of ${\mathrm K}^{0}_{S}$, $\Lambda$ and $\bar{\Lambda}$ measured by ALICE are obtained in the inelastic event (INEL) category and are observed to be 20\% lower than the rates measured by CMS for non-single-diffractive (NSD) events \cite{CMS_str_pp}. The INEL and NSD cross-section ratio $\sigma_{\mathrm{NSD}}/\sigma_{\mathrm{INEL}}$ has been measured to be of about 80\% \cite{diffractionpaper}, so this difference is consistent with the expectation that there is a negligible strange particle production at mid-rapidity in single-diffractive events. 

We compare the measured transverse momentum spectra to PYTHIA Perugia-2011 predictions \cite{skands,newtunes}. This tune is chosen because it provides the best description of multi-strange baryon production \cite{MSPP} while simultaneously also being tuned to charged particle multiplicities measured in pp collisions at 7 TeV. It can be seen in Fig.~\ref{ActualSpectra} that, while there is relative agreement at a $\ensuremath{p_{\mathrm{T}}}$ higher than 6~{\mbox{\rm GeV$\kern-0.15em /\kern-0.12em c$}}, the soft part of the spectrum shows some disagreement, albeit less than that observed for the multi-strange baryons~\cite{MSPP}.

\section{Measurements in Heavy-Ion Collisions}
\begin{figure}
\includegraphics[width=3.2in]{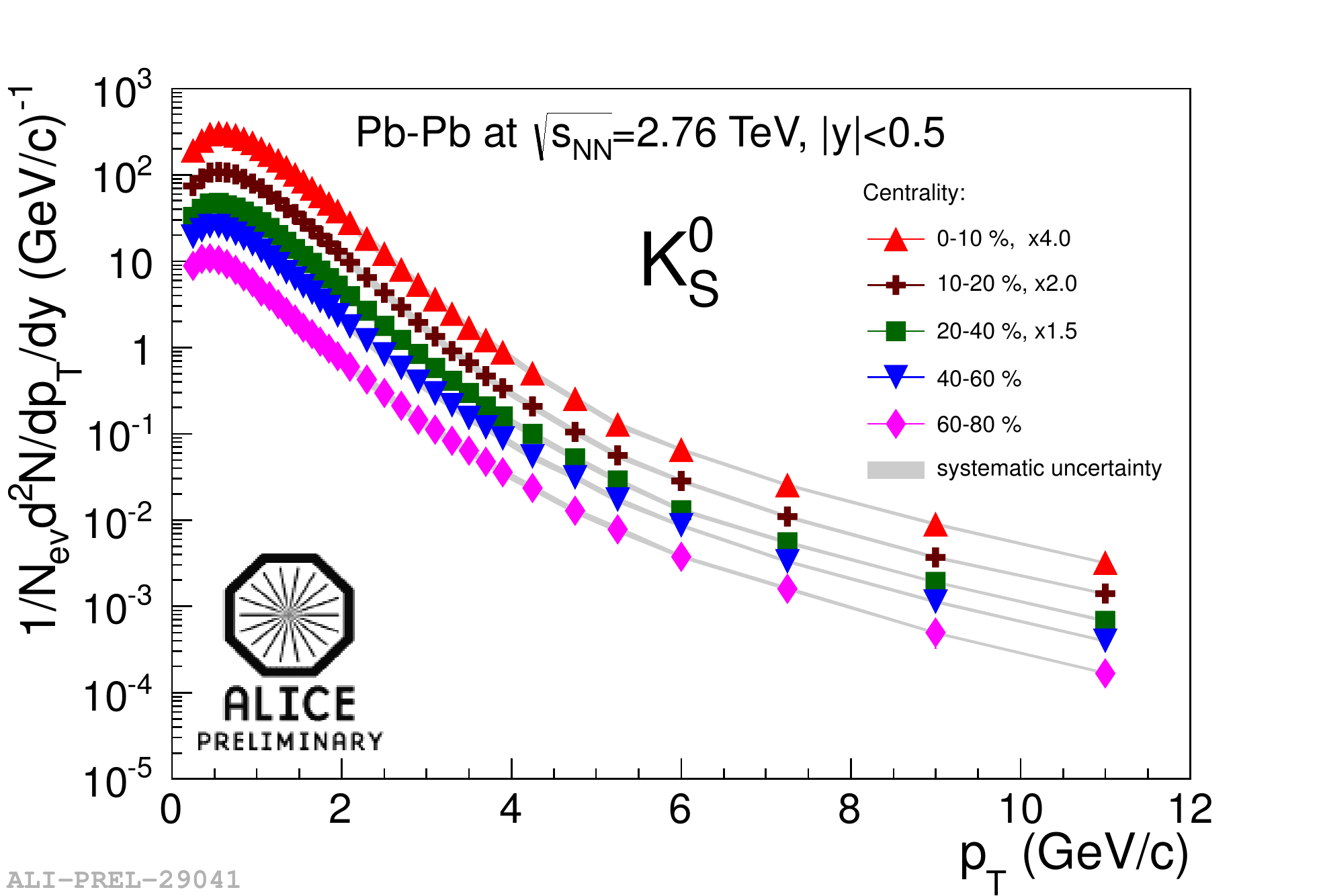}
\includegraphics[width=3.2in]{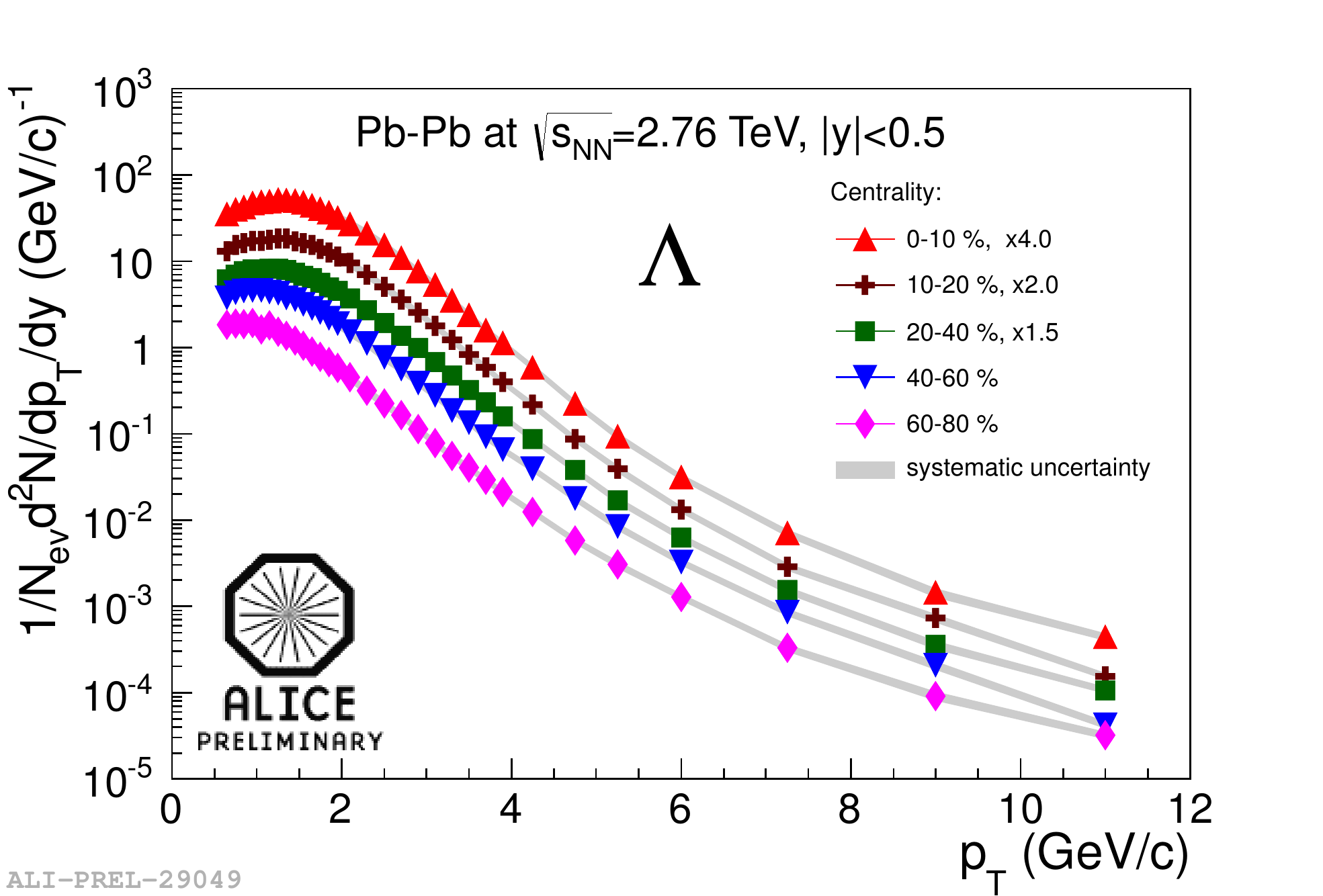}
\caption{\label{SpectraPbPb}${\ensuremath{p_{\mathrm{T}}}}$ spectra of ${\mathrm K}^{0}_{S}$ (left) and $\Lambda$ (right) in Pb-Pb collisions for 5 centrality intervals.}
\end{figure}

Figure \ref{SpectraPbPb} shows the ${\mathrm K}^{0}_{S}$ and $\Lambda$ transverse momentum spectra in Pb-Pb collisions at \twosevensix. $\bar{\Lambda}$ spectra are not drawn, however the $\bar{\Lambda}/\Lambda$ ratio is compatible with unity. We observe that the ratio $\Lambda/\pi$ is constant with centrality (not shown), which means chemistry does not significantly change with centrality. A more thorough evaluation of the hadrochemistry in heavy-ion collisions is currently under development by combining these measurements and others, such as of other, non-strange particles and resonances.

\begin{figure}
\centerline{\includegraphics[width=3.75in]{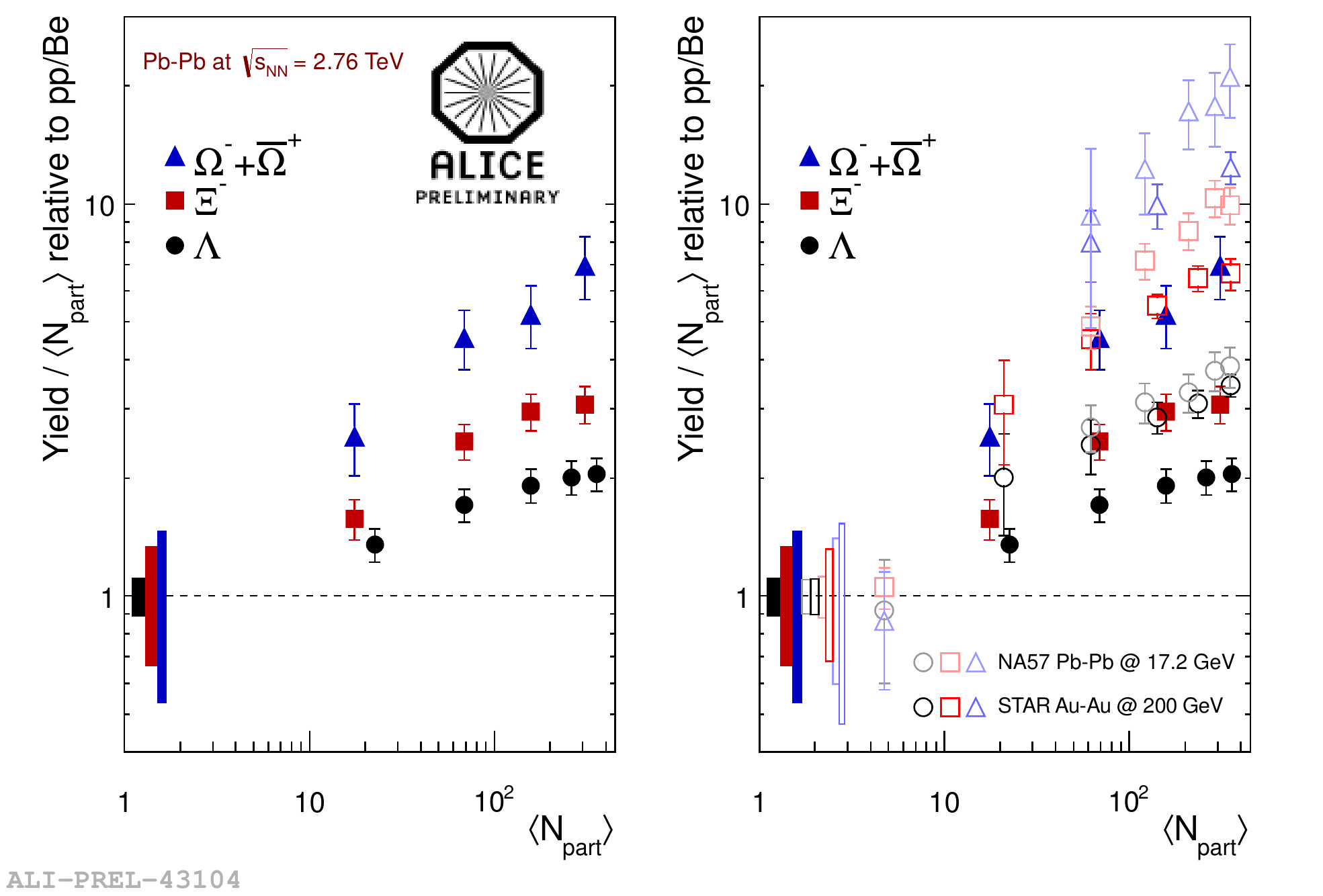}}
\caption{\label{StrangenessEnhancement}  Strangeness enhancement for $\Lambda$, $\Xi^{-}$ and $\Omega^{-}$ shown with $\langle N_{\mathrm{part}}\rangle$ scaling. Right panel: comparisons to lower energy measurements from NA57 and STAR.}
\end{figure}

We have also measured transverse momentum spectra for $\Xi^{-}$, $\bar{\Xi}^{+}$, $\Omega^{-}$ and $\bar{\Omega}^{-}$, thus enabling the study of strangeness enhancement for $|S|=2$ and $|S|=3$ cases. This can be seen in Fig.~\ref{StrangenessEnhancement}, where $N_{\mathrm{part}}$ scaling is used to account for the system size in nuclear collisions in the pp comparison. In this plot, we see an increasing enhancement with both increasing strangeness content as well as with increasing centrality. When compared to lower energy measurements done at SPS \cite{StrEnhNA57_1,StrEnhNA57_2} and at RHIC \cite{StrEnhSTAR}, the enhancement is seen to decrease with increasing beam energy, continuing the trend first observed at the SPS. However, since the $\Lambda/\pi$ ratio has been shown to be constant with centrality and, additionally, the charged particle multiplicity does not scale with $N_{\mathrm{part}}$ \cite{ChargedParticleMultDensityLeadLead}, other alternatives to the $N_{\mathrm{part}}$ scaling are currently being studied to quantify and factor out the effect of a general increase in particle multiplicity and that of an actual strangeness enhancement.

\section*{References}

\bibliographystyle{iopart-num}
\bibliography{hq2012-v01.bib}

\providecommand{\newblock}{}
\begin{thebibliography}{10}
\expandafter\ifx\csname url\endcsname\relax
  \def\url#1{{\tt #1}}\fi
\expandafter\ifx\csname urlprefix\endcsname\relax\def\urlprefix{URL }\fi
\providecommand{\eprint}[2][]{\url{#2}}

\bibitem{RafelskyStranEnh}
{Rafelski} J and {M{\"u}ller} B 1982 {\em Phys. Rev. Lett.\/} {\bf 48}
  1066--1069

\bibitem{pythia6.4}
Sjostrand T, Mrenna S and Skands P~Z 2006 {\em JHEP\/} {\bf 05} 026

\bibitem{HijingRef}
Wang X~N and Gyulassy M 1991 {\em Phys. Rev. D\/} {\bf 44}(11) 3501--3516

\bibitem{Geant3Ref}
Brun R {\em et~al.\/} 1985 {\em CERN Data Handling Division DD/EE/841\/}

\bibitem{Geant3Ref2}
Brun R {\em et~al.\/} 1994 {\em CERN Program Library LongWrite- up, W5013\/}

\bibitem{ALICE_str_pp}
Aamodt K {\em et~al.\/} 2011 {\em Eur. Phys. J. C\/} {\bf 71} 1594

\bibitem{CMS_str_pp}
Khachatryan V {\em et~al.\/} (CMS) 2011 {\em JHEP\/} {\bf 2011}(5) 1--40

\bibitem{diffractionpaper}
Abelev B {\em et~al.\/} (ALICE Collaboration) 2012  (\textit{Preprint}
  \eprint{arXiv:1208.4968v1})

\bibitem{skands}
Skands P~Z 2010 {\em Phys. Rev.\/} {\bf D82} 074018

\bibitem{newtunes}
Field R 2010  (\textit{Preprint} \eprint{arXiv:1010.3558v1})

\bibitem{MSPP}
Abelev B {\em et~al.\/} 2012 {\em Phys. Lett. B\/} {\bf 712} 309 -- 318

\bibitem{StrEnhNA57_1}
Antinori F {\em et~al.\/} (NA57 Collaboration) 2006 {\em J. Phys. G\/} {\bf 32}
  427

\bibitem{StrEnhNA57_2}
Antinori F {\em et~al.\/} (NA57 Collaboration) 2010 {\em J. Phys. G\/} {\bf 37}
  045105

\bibitem{StrEnhSTAR}
Abelev B~I {\em et~al.\/} (STAR Collaboration) 2008 {\em Phys. Rev. C\/} {\bf
  77}(4) 044908

\bibitem{ChargedParticleMultDensityLeadLead}
Aamodt K {\em et~al.\/} (ALICE Collaboration) 2011 {\em Phys. Rev. Lett.\/}
  {\bf 106}(3) 032301

\end{thebibliography}
\end{document}